\documentclass[usenatbib]{mn2e}
\usepackage{psfig}

\title{The \emph{K} band galaxy luminosity functions of three massive 
high redshift clusters of galaxies.}

\author[S.C. Ellis and L.R. Jones]
  {S.C.~Ellis$^{1,2}$\thanks{E-mail: sce@aaoepp.aao.gov.au} and L.R.~Jones$^{2}$ \\
$^{1}$Anglo-Australian Observatory, PO Box 296, Epping, NSW 2121, Australia.\\   
 $^{2}$School of Physics and
Astronomy, University of Birmingham, Birmingham, B15 2TT, UK. \\}

\date{Accepted .....................; Received .....................;
in original form .......................}  

\begin{document} 

\maketitle 

\begin{abstract} 
\emph{K} band observations of the galaxy populations of three high redshift ($z=0.8$--$1.0$), X-ray selected, massive clusters are presented.  The observations reach a depth of $K \simeq 21.5$, corresponding to $K^{*}+3.5$ mag.  The evolution of the galaxy properties are discussed in terms of their \emph{K} band luminosity functions and the \emph{K} band Hubble diagram of brightest cluster galaxies.

The bulk of the galaxy luminosities, as characterised by the parameter
$K^{*}$ from the \citet{sch76} function, are found to be consistent
with passive evolution with a redshift of formation of $z_{f}\approx
1.5$--2.  This is consistent with observations of other high redshift
clusters, but may be in disagreement with galaxies in the field at
similar redshifts.  
A good match to the shape of the Coma cluster luminosity function is
found by simply dimming the high redshift luminosity function by an
amount consistent with passive evolution.
The evolution of the  cumulative fraction of $K$ band light as a function of luminosity 
shows no evidence of merger activity in the brighter galaxies. 

The evolution of the brightest cluster galaxies (BCGs) is tested by
their \emph{K} band Hubble diagram and by the fraction of \emph{K}
band cluster light in the BCGs.  The evolution observed  is consistent
with recent previous observations although the scatter in the Hubble diagram
allows for a range of evolutionary histories. 
The fraction of cluster light contained in the BCGs is not smaller
than that in Coma, suggesting that they are already very massive with
no need to 
hypothesise significant mergers in their futures.

\end{abstract}


\section{Introduction.}

The evolutionary history of galaxies in clusters remains a subject of debate.  
The two most common explanations of massive early-type galaxy formation and evolution are those of monolithic collapse (e.g.\ \citealt{egg62}) and hierarchical merging (e.g.\ \citealt{col94}).  In the monolithic collapse scenario all the galaxies (and the stars therein) are formed in a single burst and subsequently evolve passively (along the main sequence) with no further star formation.  Such a model will result in a very homogeneous population of galaxies since their ages and metallicities will all be close to identical, with a small degree of scatter reflecting the variations in initial mass function at formation.   There is a significant body of observational evidence that luminous early type cluster galaxies are indeed remarkably homogeneous.  For example, the observed tightness of the colour-magnitude relation (e.g.\ \citealt{vis77}, \citealt{bow92b}) is naturally explained by such a homogeneous population.

In the merging model galaxies form by a series of mergers within a hierarchical model of structure formation (e.g.\ \protect\citealt{kau93}).  The hierarchical scenario presents a radically different evolution than in the monolithic collapse picture.  Bursts of star formation, related to mergers, may occur and a more gradual increase in the number of massive galaxies in a cluster would then occur as small galaxies merge to form larger ones.  



Although the ages of the majority of the stars in galaxies in both
scenarios are similar (in order to explain the tightness of the
colour-magnitude relation and the evolution of the fundamental plane),
the mass as a function of look-back time will be quite different,
i.e.\ the epoch of the assembly of massive galaxies is not the same as
the epoch of star formation in the merging case.  The hierarchical
model will exhibit negative evolution of mass as a function of
redshift (i.e.\ at higher redshifts the massive galaxies will be less massive
on average), whereas in the monolithic collapse model galaxies will
have a constant mass with redshift.  The negative evolution expected
from hierarchical models will be most apparent in the most massive
galaxies, which are predicted to have assembled more recently.
The redshift of assembly of massive galaxies is, however, poorly known and is probably dependent on environment.



If one is interested in the stellar mass of the galaxies, as opposed
to their star formation rate, then the \emph{K} band is a good choice
in which to make observations since the light at such wavelengths
originates mainly from the longer lived stars of the main sequence
(see e.g.\ figure~1 of \protect\citealt{kau98b}), and has the added
advantages that the k-correction differences between galaxies of
different spectral types are small in \emph{K} and the Galactic
extinction by dust is small.  Thus computing \emph{K} band luminosity
functions for clusters of galaxies should give a useful measure of the
mass distribution of galaxies within clusters (see
section~\ref{sec:klf} and \protect\citealt{tre98b}).  




Brightest cluster galaxies (BCGs) are known to have a very limited
variation in absolute magnitude which historically led to their
nomination as a  candidate for a cosmological standard candle with
which to directly measure the curvature of the Universe
(\protect\citealt{san72a}, \protect\citealt{san72b}.)  Controversy
over their suitability as standard candles is believed to be due to
environment, with BCGs in high $L_{\mathrm{X}}$    clusters exhibiting
much less scatter than those in low  $L_{\mathrm{X}}$ systems
(\protect\citealt{bro02}, \protect\citealt{bur00}).  However the
evolution of BCG properties  with redshift is now the primary  interest of research on BCGs, since there is much evidence that they are a special case in the evolution of galaxies within clusters.  For instance, it is known that BCGs often do not follow the same luminosity function as other galaxies in clusters (\protect\citealt{sch76}, \protect\citealt{dre78}, \protect\citealt{bha85}), most probably due to the fact that they have a peculiar formation history.  Therefore the evolution of BCGs can provide a different and complementary study of galaxy formation theories compared with the general cluster population.  The \emph{K} band Hubble diagram of BCGs provides an efficient measure of the evolution of BCGs, and we present our results of this in section~\ref{sec:bcg}.

Throughout this paper we have used a cosmology of  $H_{0}=70$ km s$^{-1}$ Mpc$^{-1}$ in a flat universe with $\Omega_{\mathrm{M}}=0.3$ and $\Omega_{\Lambda}=0.7$, except where stated.

\section{Data.}
\label{data}

\subsection{The Clusters.}
\label{sec:gals}

We present a study of three of the most massive ($\sim 10^{15}$M$\odot$ \protect\citealt{mau03a}, \protect\citealt{mau03b}), high redshift clusters known.  They are thus ideal probes of galaxy evolution
within clusters. 
Such massive clusters at high redshift are very rare and we have an unusual opportunity to study the galaxy populations of rich, distant clusters and compare results with local massive clusters such as Coma.  The high redshift of the clusters should make any evolution in the galaxy populations easier to observe.
Two of the clusters (ClJ1226 and ClJ1415)  appear relaxed 
based on their  X-ray morphologies and one of them (ClJ0152) is probably in a state of merging.  
Thus we also have a small selection of different environments. 
All three clusters are X-ray selected and thus should be relatively
free from projection effects. The galaxy properties should also be largely
independent of the selection process. 
They were all discovered in the Wide
Angle \emph{ROSAT} Pointed Survey (WARPS, \protect\citealt{sch97},
\protect\citealt{jon98a}, \protect\citealt{per02}).
Details are given below on the X-ray properties of each cluster. 

\subsubsection{ClJ0152}
\label{sec:0152}

The cluster ClJ0152.7-1357 (\protect\citealt{ebe00}, \protect\citealt{del00}, \protect\citealt{mau03a})
is at a redshift of $z=0.833$ and has a bolometric X-ray luminosity of $1.6 \pm 0.2 \times 10^{45}$ ergs s$^{-1}$. It is
composed of two major subclumps which are probably gravitationally
bound (\protect\citealt{mau03a}). The projected separation of the
subclumps is $\approx$1.5 arcmin (or 720 kpc), suggesting that they may be in the early stage of
a very massive merger event. 
Individually the subclumps have X-ray temperatures and luminosities of $L_{\mathrm{X}}=1.0 \pm 0.2 \times 10^{45}$ ergs s$^{-1}$, $T_{\mathrm{X}}=5.5^{+0.9}_{-0.8}$ keV and $L_{\mathrm{X}}=5.8^{+1.1}_{-0.9} \times 10^{44}$ ergs s$^{-1}$, $T_{\mathrm{X}}=5.2^{+1.1}_{-0.9}$ keV respectively.  Thus, each clump considered individually is still a massive cluster with $M_{\mathrm{total}}\approx 6 \times 10^{14}$ M$_{\odot}$ and $M_{\mathrm{total}}\approx 5 \times 10^{14}$ M$_{\odot}$ for the northern and southern clumps respectively (see \protect\citealt{mau03a}).

\subsubsection{ClJ1226}
\begin{figure}
\psfig{file=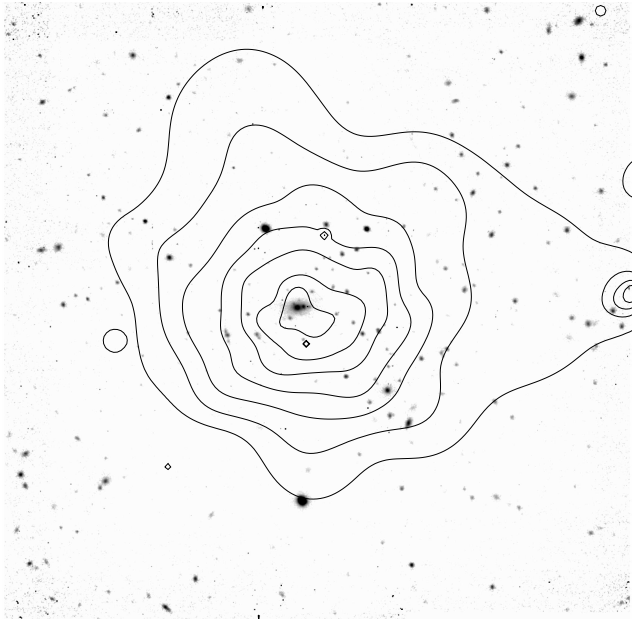,width=1\columnwidth}
\caption{\emph{UKIRT} UFTI $K$ band image of the $z=0.89$ cluster
  ClJ1226 with adaptively smoothed \emph{XMM} X-ray contours from
  Maughan et al. (2003b). The image is $\sim 2.9$ arcmin square.}
\label{fig:1226}
\end{figure}

ClJ1226.9+3332 (\protect\citealt{ebe01}, \protect\citealt{mau03b},
\protect\citealt{cag01}) is at a redshift $z=0.888$.  It has a bolometric
X-ray luminosity of $L_{\mathrm{X}}=5.3^{+0.2}_{-0.2} \times 10^{45}$
ergs s$^{-1}$, an X-ray temperature of
$T_{\mathrm{X}}=11.5^{+1.1}_{-0.9}$ keV and
$M_{\mathrm{total}}=1.4^{+0.2}_{-0.2} \times 10^{15}$ M$_{\odot}$
(\protect\citealt{mau03b}), making it one of the most massive clusters
known at high redshifts. The cluster appears remarkably relaxed in
X-rays indicating little dynamic activity.  The $K$ band image with
adaptively smoothed  \emph{XMM} X-ray contours (from Maughan et al 2003b) is shown in figure~\ref{fig:1226}.

\subsubsection{ClJ1415}
ClJ1415.1+3612 is at the highest redshift of the three clusters with
$z=1.03$ (\protect\citealt{per02}).  Although at present we have only {\sc
  ROSAT} X-ray data for the cluster so sub-structure may not be resolved, it exhibits smooth X-ray contours suggesting it may be dynamically relaxed. It has a bolometric X-ray 
luminosity of $L_{\mathrm{X}} = 2 \times 10^{45}$ ergs s$^{-1}$, and is thus likely to be a very massive cluster.

\subsection{Observations.}
\label{sec:obs}

All three clusters were observed at the 3.8m United Kingdom Infrared Telescope ({\sc UKIRT}) on Mauna Kea, Hawaii using the {\sc UFTI} camera, a $1024 \times 1024$ pixel HgCdTe array, with a pixel size of 0.091 arcseconds.  Observations were made with the K98 and J98 filters (50\% cut-offs 1.17--1.33$\mu$m and 2.03--2.37$\mu$m respectively) with an exposure time of 66s per frame (or 60s per frame for a
minority of observations made in service time) and the seeing was
typically $\approx 0.5$ arcseconds in $K$.  The total exposure times in
$K$ are listed in table~\ref{tab:calibration} along with the average
seeing for each cluster. A spatial dithering of 15 arcsec was employed. Since the field of view of 90x90 arcsec covered only a fraction of each cluster, mosaics of different sizes for each cluster were used. 
On each night standard stars (\protect\citealt{haw01}) were observed in order to calibrate the zero-point, air mass coefficient, and colour coefficient.  Some nights had patchy cloud cover, and thus yielded inaccurate photometric calibration.  Such nights were deemed to be non-photometric,
and a calibration was obtained via observations of the same fields on photometric nights.
Non-photometric data were then weighted according to their shorter effective exposure times.

Observations were also made of one or more offset fields in the vicinity of each cluster on the sky, chosen as a reasonably `blank' part of a digitised sky survey image (since at these redshifts, the clusters are
also blank on   digitised sky survey images).  The purpose of these
fields was to provide a means of estimating the contamination from
foreground and background galaxies when constructing luminosity
functions. The offset fields were typically 4 arcmin (1.9 Mpc) from
the clusters, beyond the virial radii.

Observations of ClJ0152 were made in service time, on the nights of 15th and 20th October 2000, 24th December 2000, 28th December 2000 and 16th January 2001. 
The northern and southern clumps will be referred to as field A and field B. 
The nights of 16th January, 24th December and 28th December were found to be non-photometric and were treated accordingly.  The areas observed were $1'50'' \times 1'39''$ and $1'51'' \times 1'50''$ for fields A and B respectively, corresponding to a physical size of $836 \times 752$kpc and $843 \times 836$ kpc.

ClJ1226.9+332 was observed on the 18th, 19th and 20th April 2001.  The night of the 18th had patchy cloud cover and was deemed non-photometric.
The observations covered an area of $2'47'' \times 3'1''$, corresponding to a physical size of $1.3 \times 1.4$ Mpc.

Cl1415 was observed on the 10th, 11th, 12th April 2002.  All nights were photometric with an average seeing in \emph{K} of 0.52''.  The observations covered an area of $2'2'' \times 2'44''$, corresponding to an area of $1.0 \times 1.3$ Mpc. For this cluster, with the highest redshift, two offset fields were observed in order to reduce the uncertainties in the field subtraction. For the other clusters,
one offset field was observed.

\subsection{Data Reduction.}
\label{data:reduction}

Each observation of a particular field consisted of a number of dithered
frames of the object plus a dark frame
containing only the instrumental background.  The data were 
dark-subtracted, flat-fielded and combined 
using standard {\sc iraf} procedures. A bad-pixel mask was applied to exclude any defective pixels.
Flat-field frames were produced from the object frames,
using the median value of each pixel
after applying a 4$\sigma$ clipping and scaling each frame by its modal value to account for
sky brightness variations. 

Each frame had its background sky value subtracted, using a $3\sigma$-clipped mean as an estimate of the sky value. The relative offsets of each frame were then measured
using bright stars in the field of view. Finally the frames were shifted and combined using the mean value of each pixel, again
after applying a 4$\sigma$ clipping around the median value.

\section{Photometry.}
\label{photometry}

\subsection{Standard stars.}
\label{standards}

The photometry of the galaxies was calibrated using standard stars 
observed on the same nights as the cluster fields.   The true magnitude of any celestial body is
taken to be
\begin{equation}
m_{{\rm true}} = ZP - 2.5 {\rm log} (CR) + A{\rm sec}z + B(J-K)_{{\rm true}}
\label{eqn:mag}
\end{equation}

where $ZP$ is the
zero-point magnitude, $CR$ is the count rate, $A$ is the coefficient of extinction per unit airmass,
$z$ is the zenith angle (sec$z$ therefore being the airmass) and $B$ is the
colour coefficient necessary due to differences in the response of the combination of
camera and filters used and the standard star system. 







Values of $ZP$, $A$ and $B$ were determined from observations of standard stars taken on each night.  Standard stars were selected from the \emph{UKIRT} faint standards list (\protect\citealt{haw01}).  Typically each night 6 stars of varying colours were observed between 1 and 3 times each at differing airmasses.

Whilst the value of $A$ and $ZP$ may vary from night to night, the value of the colour coefficient 
$B$ should be almost constant. Thus a single value of $B$ was measured from all the photometric nights
combined. The value we obtained is consistent with zero, as expected from the filter design (\protect\citealt{tok02}). An independent value of $ZP$ was measured for each night, and for most nights,
an independent value of $A$ was measured. Where this was not possible a value of $A$   
from other nights was used and checked for consistency.  All were found to be consistent.
In any case the low extinction in the K band and the low airmasses at which most observations
were performed combine to make this a relatively small correction ($<$0.2 mag).

The 
parameters are listed in table~\ref{tab:calibration}.
Note that the final photometric error, as estimated from the scatter
in the standard star measurements, is less than 0.1 mag.
Corrections due to Galactic extinction  are negligible ($\approx$0.007 mag) and were not
applied.

\begin{table*}
\caption{Calibration parameters for $K$ band.}
\begin{tabular}{|l|l|l|l|l|}
& ClJ0152A & ClJ0152B & ClJ1226 & ClJ1415 \\ \hline
$A$ & $-0.150 \pm 0.110$ &$-0.150 \pm 0.110$ & $-0.178 \pm 0.052$ & $-0.088$ \\ 
$B$ & \multicolumn{4}{c|} {$-0.0043 \pm 0.094$} \\
$ZP$ &22.248$\pm 0.018$& 22.248$\pm 0.018$ & $22.367\pm 0.008$ & 22.422$\pm 0.044$ \\
Mean sec$z$ & 1.445 & 1.403 &1.210 & 1.160 \\
Exposure/ s & 3882 & 3942 & 23946 & 17820 \\
Seeing/ arcsec & 0.53 & 0.53 & 0.57 & 0.52 \\
\label{tab:calibration}
\end{tabular}
\end{table*}

\subsection{Galaxies.}
\label{galaxies}

The
{\sc SExtractor} software of \citet{ber96} was used to search for
objects in each field.  To detect objects a threshold value per pixel
must be chosen along with a minimum number of connected pixels.
Because the final images are constructed from a jittered pattern of
images, the depth of observation varied across the final image, being
at its deepest in the centres and shallowest at the edges.  Therefore
detection of objects was done in two regions for each image, using
different detection parameters, in order to reach as deep as possible
in the centre of the image whilst avoiding spurious detections at the
edges of the image.  Typically the minimum significance of the detections were
3.5$\sigma$ or 4$\sigma$ in the centres of the images and $7\sigma$ at
the edges, where $\sigma$ is the background RMS determined from counts
over the whole image (and is thus an overestimate of $\sigma$ in the
central region and an underestimate in the outer region, although
 this is not the case for the central region of ClJ1415, when $\sigma$ was determined from the central region alone). Reliability of the object detection, and in particular 
the handling of overlapping objects, was checked by eye.  Faint pixels surrounding deblended objects are assigned to one of the sub-objects with a probability based on the expected contribution at that pixel from each of the deblended objects (see \protect\citealt{ber96}).

 Counts were measured using an adaptive aperture based
on Kron's algorithm (\citealt{kro80}), and also in a circular aperture with radius chosen to maximise the signal to noise.  For each object a value of
stellarity was also measured (see \citealt{ber96}).

To determine the magnitudes of the objects equation~\ref{eqn:mag} was used
with the following complication.  It is unknown to start with what the
true colours of the objects are, therefore the  magnitudes cannot
be determined.  To circumvent this problem an approximate magnitude was
measured in the \emph{J} and \emph{K} bands neglecting the colour term in equation~\ref{eqn:mag}.  This allowed
an approximate colour to be determined.  The approximate colour is then
multiplied by a correction factor previously determined using the same
technique to measure the approximate colours of standard stars and their
true colours.  The average value of
$\frac{(J-K)_{{\rm true}}}{(J-K)_{{\rm approx}}}$ was 0.982.

Note that for the determination of colours, magnitudes derived from fixed,
circular apertures were used, whereas the adaptive aperture magnitudes were 
used to derive pseudo-total magnitudes.  The use of adaptive aperture magnitudes using {\sc SExtractor}'s Kron radius should avoid the problem described by \citet{and02} of underestimating fluxes for galaxies with low central surface brightness. The overall reliability of the photometry was checked
by comparing the field galaxy counts derived from the offset fields with deeper published 
results (see fig.~\ref{fig:fieldcounts}). There is generally good agreement down to our limiting magnitudes (see below).


Star--galaxy discrimination was determined using {\sc SExtractor's} stellarity parameter.  A cut-off of 0.8 was selected to delimit the two classes of objects with those objects with values greater than 0.8 being excluded as stars.  This value was confirmed by examination of the radial profiles of detected objects.  It was found that objects with a stellarity greater than 0.8 had an almost constant Gaussian FWHM close to the value of the seeing, whereas objects with stellarity less than 0.8 had more extended profiles.  The results are insensitive to the precise value of this cut-off as there were very few stars in each field.


  
\section{Luminosity functions.}
\label{sec:klf}

\emph{K} band luminosity functions (KLFs) were calculated separately for each cluster of galaxies.  This was done in the following way. Firstly,  some of the brightest galaxies have been spectroscopically
confirmed as cluster members, so where there are no other bright galaxies in the field,
no subtraction is necessary.  At fainter magnitudes, the number of galaxies, $N_{\mathrm{cl}}$, in a bin of width $\Delta M$ was counted, with  $\Delta M=1$ for ClJ0152 and ClJ1415 and $\Delta M=0.75$ for ClJ1226.  The contamination from foreground and background galaxies was estimated by counting the number of galaxies, $N_{\mathrm{back}}$ in the offset field(s) associated with the cluster.  Thus the number of galaxies in each bin is given by,

\begin{equation}
\label{eqn:nbin}
N_{\mathrm{bin}} = N_{\mathrm{cl}}-N_{\mathrm{back}} \times \left( \frac{A_{\mathrm{cl}}}{A_{\mathrm{back}}} \right)
\end{equation}

where $A_{\mathrm{cl}}$ is the area of the cluster field and $A_{\mathrm{back}}$ is the area of the offset field.  Note that because two different detection thresholds have been used for each field the faintest bins are from a smaller area than the brighter bins (since only a small area was deep enough to detect objects of this magnitude).  Therefore the detections from the smaller area have been scaled by an appropriate factor to bring them into line with the detections from the total area.

The corresponding error on the number of galaxies in each bin is found by summing in quadrature the error on $N_{\mathrm{cl}}$ and the error on $N_{\mathrm{back}}$,

\begin{equation}
\label{eqn:errorn_bin}
\sigma_{\mathrm{Nbin}} = \sqrt{\sigma_{\mathrm{Ncl}}^{2} + \sigma_{\mathrm{Nback}}^{2}}
\end{equation}

where $\sigma_{\mathrm{Ncl}}$ is the Poissonian error (\protect\citealt{geh86}).  The background error, $\sigma_{\mathrm{Nback}}$, has a Poissonian term and a term taking into account the clustering properties of galaxies, viz.,

\begin{equation}
\label{eqn:error_back}
\sigma_{\mathrm{Nback}} = \sqrt{N_{\mathrm{back}}} \times \sqrt{1+\frac{2\pi\it{N}A_{\omega}\theta_{\mathrm{c}}^{2-\delta}}{2-\delta}}
\end{equation}

where $\it{N}$ is the number density of galaxies in the bin and $\theta_{\mathrm{c}}$ is the angular radius such that $\Omega=\pi\theta_{\mathrm{c}}^{2}$ where $\Omega$ is the solid angle of the background field.  The parameters $A_{\omega}$ and $\delta$ describe the angular correlation function of galaxies such that,

\begin{equation}
\label{eqn:ang_corr}
\omega(\theta)=A_{\omega}\theta^{\delta}.
\end{equation}

(\protect\citealt{pee80}, pg.175).  

The value of $\delta=-0.8$ was used and $A_{\omega}$ was found using the formula

\begin{equation}
\label{eqn:aomega}
\mathrm{log}_{10}A_{\omega}=7.677 - 0.3297r_{\mathrm{lim}}
\end{equation}

derived by fitting to the data in table~1 of \citet{bra95}.  Following \citet{djo95} we used $r_{\mathrm{lim}}-K_{\mathrm{lim}}\sim 3$.  

\begin{figure}
\psfig{file=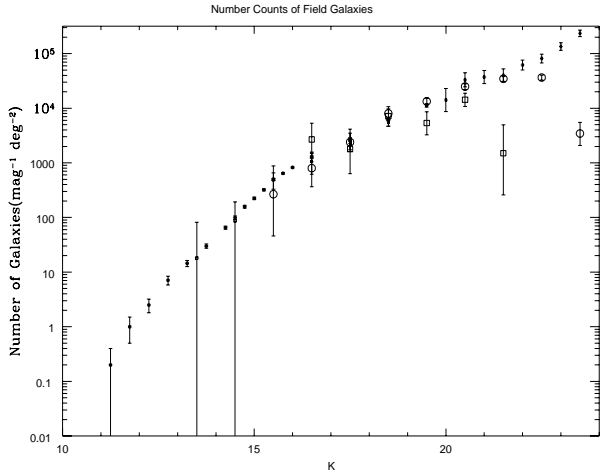,width=8.5cm,height=6.5cm,angle=270}
\caption{Number density of field counts as a function of magnitude.
  Open circles are from the combined offset fields of ClJ1226 and
  ClJ1415, open squares are from the offset field of ClJ0152, filled
  circles from \protect\citet{hua97}, filled squares from \protect\citet{sar97} and stars from \protect\citet{djo95}.}
\label{fig:fieldcounts}
\end{figure}

Our limiting \emph{K} magnitude was determined by plotting the number density of galaxies from the offset fields as a function of magnitude and determining the drop-off point when compared to large area field surveys.  Figure~\ref{fig:fieldcounts} shows the number density of galaxies in our four offset fields. ClJ1226 and ClJ1415 were found to be complete to $K_{\mathrm{lim}} = 21.5$ and ClJ0152 was found to be complete to $K_{\mathrm{lim}} = 20.0$.

No corrections have been made for gravitational lensing of the
background galaxies by the cluster dark matter. This effect
magnifies the fluxes of background galaxies. Trentham \& Mobasher
(1998) show that the net increase in  the number of galaxies to be
subtracted is small ($<$5\%) at the redshifts and magnitudes
considered here (and for an equally massive cluster), because a large
fraction of the field galaxies along the line of sight are in the
foreground.  This correction is significantly less than the
statistical errors.

\begin{figure}
\psfig{file=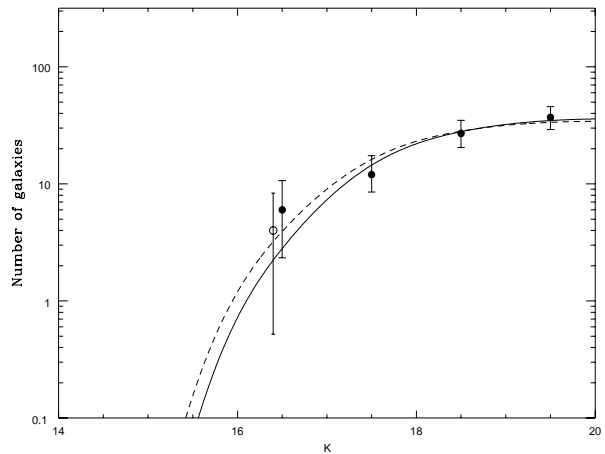,width=8.5cm,height=6.5cm,angle=270}
\caption{Binned \emph{K} band luminosity function of ClJ0152 (z=0.83).  Also shown is the best fitting Schechter function.  The solid line excludes the BCGs (one from each sub-clump)and the dashed line includes the BCGs. The open symbol (offset slightly for clarity) shows the effect of excluding the BCGs from each sub-clump.}
\label{fig:klf0152}
\end{figure}

\begin{figure}
\psfig{file=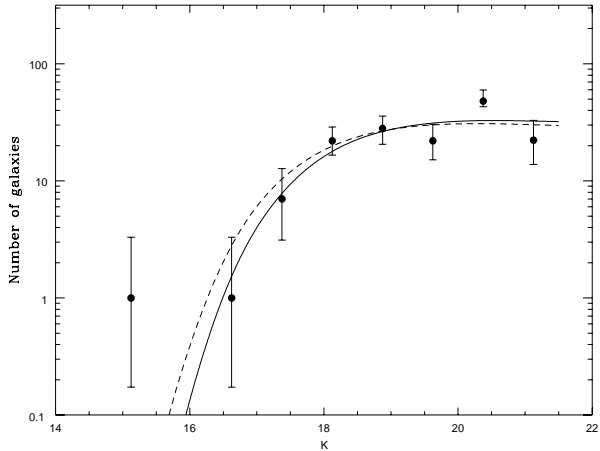,width=8.5cm,height=6.5cm,angle=270}
\caption{\emph{K} band luminosity function of ClJ1226 (z=0.89).  The best
  fitting Schechter functions including (dashed line) and excluding
  (solid line) the BCG are shown. }
\label{fig:klf1226}
\end{figure}

\begin{figure}
\psfig{file=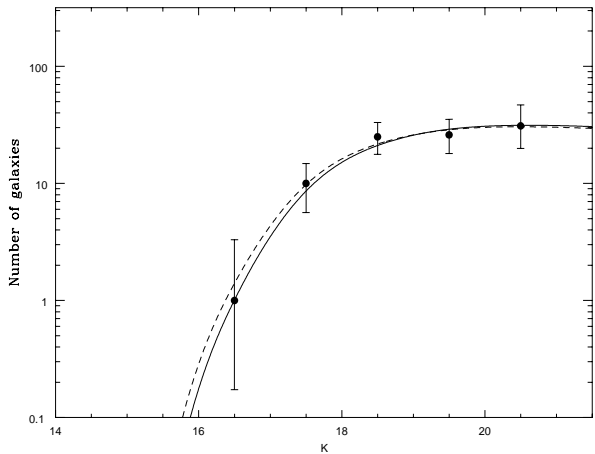,width=8.5cm,height=6.5cm,angle=270}
\caption{\emph{K} band luminosity function for ClJ1415 (z=1.03).  Schechter function fits 
are as for figure \protect\ref{fig:klf1226}.}
\label{fig:klf1415}
\end{figure}

Luminosity functions for the three clusters are shown in figure~\ref{fig:klf0152}, ~\ref{fig:klf1226} and ~\ref{fig:klf1415} along with the best fitting Schechter functions as described below.

\subsection{Fitting Schechter functions.}

The parametric \citet{sch76} luminosity function is of the form,

\begin{equation}
\label{eqn:schechter_lum}
\frac{\mathrm{d}\phi}{\mathrm{d}L}\mathrm{d}L=\phi^{*^{'}}(\frac{L}{L^{*}})^{\alpha}e^{-(\frac{L}{L^{*}})}\mathrm{d}(L/L^{*})
\end{equation}

where $L^{*}$ is the characteristic luminosity at which the function turns over to the faint end which has slope $\alpha$.  The normalisation of the function is given by $\phi^{*^{'}}$.  The corresponding function in terms of \emph{K} band magnitudes is

\begin{equation}
\label{eqn:schechter_mag}
\mathrm{d}\phi=\phi^{*}10^{0.4(K^{*}-K)^{\alpha+1}}e^{-10^{0.4(K^{*}-K)}}\mathrm{d}K
\end{equation}

where $K^{*}$ is the characteristic \emph{K} band magnitude.  Note that $\phi^{*}$ in equation~\ref{eqn:schechter_mag} is not identical to the $\phi^{*^{'}}$ in equation~\ref{eqn:schechter_lum}, but in fact differs by a factor $-0.4\mathrm{ln}10$.  

In general the Schechter function is found to be a good fit to composite cluster galaxy LFs, as in the 2dF Galaxy Redshift Survey (\citealt{dep03}).  However, for individual clusters, there is often observed a dip in the LF (eg.\ \citealt{biv95} for the Coma cluster, \citealt{yag02} for a sample of 10 clusters).  \citet{biv95} suggest that the dip is caused by a recent episode of star-formation in the brightest galaxies thereby enhancing their luminosities.  However, the presence of a dip in the $H$ band LF of Coma observed by \citet{and02} suggests that recent star-formation is unlikely to be the cause of the dip, since the $H$ band is rather insensitive to the enhanced blue emission accompanying star-formation.  Rather, \citet{and02} suggests that the dip is the result of combining the LFs of several different types (see also \citealt{bin88}).  However, \citet{and02} points out that such a dip is not seen in the near-infrared LFs of five clusters studied by \citet{tre98b}, nor in the $H$ band LF of the Coma cluster measured by \citet{dep98}.  If the dip is the result of the combination of different LFs for different morphological types, then the difference between the \citet{dep98} LF and the \citet{and02} LF suggests that the dip in the Coma LF may be dependent on which part of the cluster is being observed, and hence the particular morphological mix measured.  In the present case no obvious dips are present in any of our LFs, and given the limited resolution in magnitude caused by the statistical uncertainties, the use of any functions more complicated than a Schechter function is not justified.

Equation~\ref{eqn:schechter_mag} was fit to the KLFs of all 3 clusters using the maximum likelihood technique of \citet{cas79}.  This technique fits a Schechter function to the background subtracted, binned data in the following way.  The parameters $K^{*}$ and $\alpha$ are free parameters, and are allowed to vary between assumed limits.   For each combination of $K^{*}$ and $\alpha$ in the parameter space, a Schechter function is calculated, with the normalisation, $\phi^{*}$, set such that the predicted number of galaxies is equal to the actual number of galaxies observed within the same magnitude limits.   Thence the Cash statistic, $C$, can be computed,

\begin{equation}
\label{eqn:cash}
C=-2\sum_{i=1}^{N}(n_{i}{\rm ln}e_{i} - e_{i} - {\rm ln}n_{i}!)
\end{equation}

where $N$ is the number of bins, $n_{i}$ is the observed number of galaxies in the $i$th bin and $e_{i}$ is the expected number of galaxies in the $i$th bin resulting from the Schechter function.  The parameters for which $C$ is a minimum are deemed to be the best fitting parameters (see \citealt{cas79}).

 It was found that $\alpha$ could not be strongly constrained and therefore, following \citet{dep99} we fix at $\alpha = -0.9$ which is the value for the KLF of the Coma cluster (\protect\citealt{dep98}) and of the field (\protect\citealt{gar97}).  The characteristic magnitude, $K^{*}$, was fit as a free parameter and $\phi^{*}$ was determined by ensuring that the integral of the function down to the limiting magnitude was equal to the total number of galaxies to the same magnitude.

It has been noticed since early studies of cluster luminosity
  functions that the brightest cluster galaxies are a special class of
  object (\protect\citealt{pea69}) and that a better fit to the
  LF can often be made by excluding them from the fit
  (\protect\citealt{sch76}, \protect\citealt{dre78}).  Therefore
  Schechter functions were fit both including the BCGs and excluding
  them.  Note that in the case of ClJ0152 we have excluded two BCGs,
  one from each sub-clump of the system.  The best fitting parameters
  are given in table~\ref{tab:schechter_fits}, along with the
  $\chi^{2}$ probability (since the maximum likelihood statistic does
  not give a measure of the goodness of fit). 
  The best fitting Schechter functions, including and excluding the BCGs, are also shown in figures~\ref{fig:klf0152}, \ref{fig:klf1226} and \ref{fig:klf1415}.

\begin{table*}
\caption{Best fitting parameters of a Schechter function for each cluster with $\alpha=-0.9$.} 
\label{tab:schechter_fits}
\begin{tabular}{lllllll}
& \multicolumn{3}{c}{Including BCGs} & \multicolumn{3}{c}{Excluding BCGs} \\
& $K^{*}$ (lower limit, upper limit) & $\phi^{*}$ & Prob($\chi^{2}$) & $K^{*}$ (lower limit, upper limit) & $\phi^{*}$ & Prob($\chi^{2}$) \\ \hline
Cl0152 & 17.59 (17.26,17.90) & 47.89 & 0.57 & 17.76 (17.43,18.07) & 50.31 & 0.81\\
Cl1226 & 17.79 (17.52,18.03) & 57.37 & 0.01 & 18.04 (17.79,18.28) & 61.01 & 0.03 \\
Cl1415 & 17.96 (17.63,18.26) & 42.53 & 0.96 & 18.08 (17.75,18.38) & 43.80 & 0.78\\
Combined & 17.81 (17.51,18.07) & 44.69& 0.07 & 17.98 (17.68,18.26) & 45.59 & 0.16 \\
\end{tabular}
\end{table*}

For two of the three clusters studied here, good fits are found both
including and excluding the BCG (see table~\ref{tab:schechter_fits}). For ClJ1226, a marginal
fit is obtained excluding the BCG because of one high bin (at
$K=20.4$), but the fit is poorer when the BCG is included, a
conclusion supported by a visual inspection of figure 3. The values of
$K^*$ from each cluster are consistent with each other, but are
systematically fainter by $\approx$0.2 mag when the BCGs are
excluded. This difference is of similar size to the statistical uncertainty. For
consistency with previous measurements of Coma, in the following
section we will refer to $K^*$ values from the fits excluding BCGs. 

\subsection{Evolution of  $K^{*}$}

To evaluate the evolution in $K^{*}$, models of galaxy evolution were made using the synthetic stellar population (SSP) libraries of \citet{bru03}.  The parameters of the model are the star formation history, the initial mass function and the metallicity.  For all models considered here a single burst of star formation was assumed to occur (but at different epochs), the initial mass function was that of \citet{sal55} and the metallicity was assumed to be solar (i.e.\ $Z=2\%$).  Models were constructed for passive evolution in the following way. Spectral energy distributions (SEDs) were drawn    from the \citet{bru03} libraries for different ages.  A redshift of formation was chosen, corresponding to an age $t=0$.  Then for the adopted cosmology redshifts were calculated for the rest of the ages.  The SEDs were redshifted to the appropriate redshift for each age and multiplied by the filter transmission curve.  Thus the flux of the SSP could be calculated for each redshift.  The fluxes were converted to apparent magnitudes and scaled appropriately, e.g.\ for the evolution of $K^{*}$ the models were normalised to $K^{*}$ of the Coma cluster (see below).

No-evolution models are made by simply choosing an SED of a particular age and integrating the redshifted spectrum under the filter transmission curves.

\begin{figure}
\psfig{file=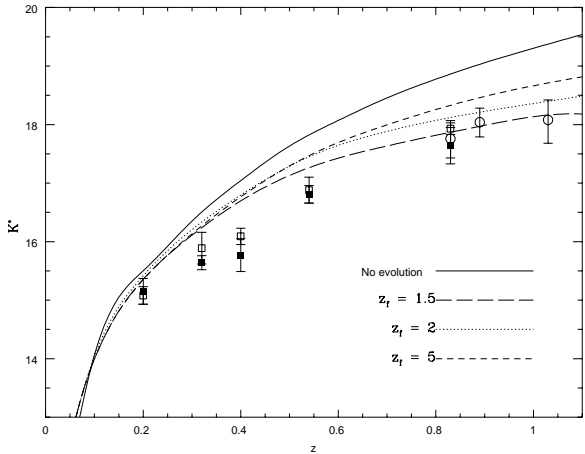,width=8.5cm,height=6.5cm,angle=270}
\caption{The evolution of $K^{*}$.  The circles are data from this paper.  The squares are from \protect\citet{dep99}, open symbols being low $L_{\mathrm{X}}$ systems and closed symbols being high $L_{\mathrm{X}}$ systems.}
\label{fig:klf_evol}
\end{figure}

Passively evolving models with redshifts of formation $z_{\rm{f}}=1.5$, $z_{\rm{f}}=2$ and $z_{\rm{f}}=5$ were made along with a no evolution model of a 10Gyr old population.  The results are shown in figure~\ref{fig:klf_evol}.  The models were scaled to the Coma cluster with $K^{*}=10.9$ at $z=0.0231$ 
(excluding BCGs, \protect\citealt{dep99}).

It is clear that the trend seen in \citet{dep99} is followed here and extended to higher redshifts. All values of $K^{*}$ are brighter than predicted by no evolution, and are generally consistent with passive evolution models.  The high redshift points seem most consistent with $z_{f}\simeq 1.5$ although the errorbars are large enough to allow a range of $z_{f}$.  
A better fit to all the data would be obtained by normalising to $K^{*}$ at $z=0.2$ (which is obtained from the mean of 3 clusters) rather than normalising to Coma.  In that case $z_{{\rm f}} \approx 2$ would be a better fit to the highest $z$ data.

Although it was not possible to constrain $\alpha$ for any individual cluster, in order to verify the choice of $\alpha=-0.9$ we have combined the counts from all three clusters into a general LF and fitted a Schechter function leaving $\alpha$ as a free parameter.

The combined LF was generated by finding a weighted average of the counts in each bin of the individual clusters where the weighting factor was the reciprocal of $\phi^{*}$. In fact the  $\phi^{*}$ values for each cluster are similar, so the weighting has little effect. The combined LFs, excluding and including BCGs, with the best fitting Schechter functions are shown in figures~\ref{fig:klfcomb} and \ref{fig:klfcomb_wbcg} respectively.  The best fitting values were $K^{*}=18.53$ and $\alpha=-0.54$, when the BCGs were not included and $\alpha=-0.94$ and $K^{*}=17.83$ when they were.  The confidence limits, excluding BCGs, are shown in figure~\ref{fig:klfcombconf} where it can be seen that the constraint on $\alpha$ is not strong, ranging from $\alpha \approx -0.15$ -- $-0.9$.  

There is a large degree of degeneracy between $K^{*}$ and $\alpha$ as is evident from the shape of the contour in figure~\ref{fig:klfcombconf}.  Consequently when fitting a Schechter function, $K^{*}$ may be misleading if the faint end slope is poorly constrained.  To quantify this effect the combined LF was fit with $\alpha=-0.9$.  The results are given in table~\ref{tab:schechter_fits}.  The result (excluding BCGs) is a brighter value of $K^{*}$, as expected from the shape of figure~\ref{fig:klfcombconf}.

Consequently, if  $\alpha\approx -0.54$ for high redshift clusters then the evolution in $K^{*}$ may not be as strong as seen in figure~\ref{fig:klf_evol}, and values of $z_f\approx 2-5$ would be more consistent with the $K^*$ measurements. However, if the last bin of the combined KLF of figure \ref{fig:klfcomb} is excluded then $\alpha=-0.99 \pm 0.4$.  Thus, it is unclear whether the measured value of $\alpha=-0.54$ is in artefact due to incompleteness in the faintest bin, and therefore we elect to use the value of $\alpha=-0.9$ for the individual fits to clusters.

\subsection{Comparison with the Coma cluster}

\begin{figure}
\psfig{file=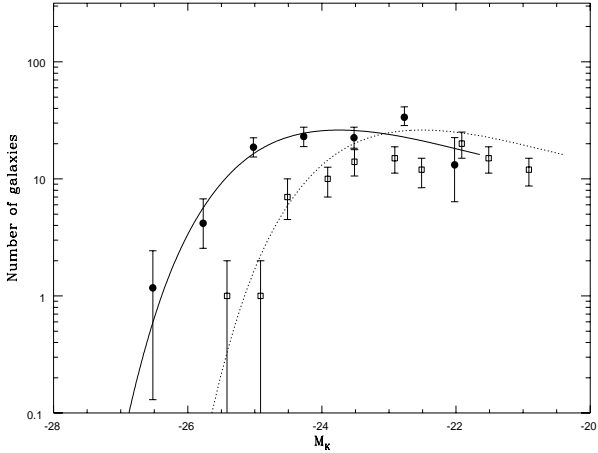,width=8.5cm,height=6.5cm,angle=270}
\caption{The combined luminosity function of all three clusters
  ($\bar(z)$=0.9) shown
  by solid circles, excluding BCGs.  Its best fitting Schechter
  function is the solid line, whilst the dotted line shows the
  expected evolution at the redshift of Coma assuming passive
  evolution and $z_{\rm{f}}=2$ .  The number of galaxies in each bin
  is a weighted average.  The Coma $K$ band luminosity function
  derived from \protect\citet{dep98} is shown by open squares. Note
  that the two LFs are plotted directly as observed, with no scaling. 
See text for details}
\label{fig:klfcomb}
\end{figure}

\begin{figure}
\psfig{file=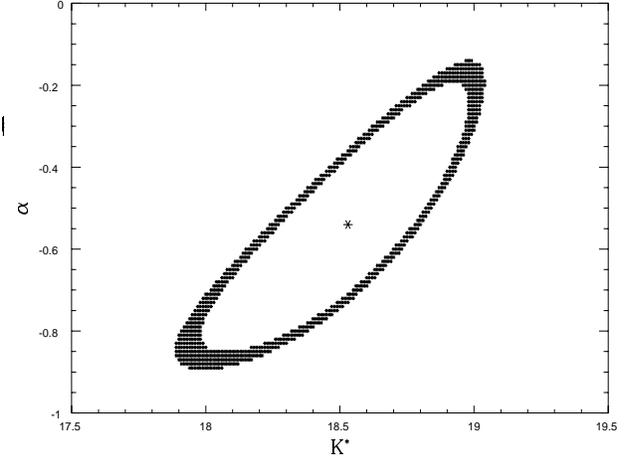,width=8.5cm,height=6.5cm,angle=270}
\caption{Confidence limits at 68 \% on the best fitting parameters for the combined luminosity function when BCGs are excluded.}
\label{fig:klfcombconf}
\end{figure}

We make a comparison with the Coma cluster because it has a similar
$L_{{\rm X}}$ (1.6x10$^{45}$ erg s$^{-1}$), $T_{{\rm X}}$ (8 keV) and mass (10$^{15}$
M$_{\odot}$) 
as the high redshift clusters, and 
it is relatively well studied at NIR wavelengths. 
The $K$ band LF of the Coma cluster shown in
figure~\ref{fig:klfcomb} was derived from the $H$ band luminosity
function of \citet{dep98} using their given value of $H-K=0.22$ and
to a magnitude limit for which all galaxies had spectroscopic
membership confirmation. Note that the two LFs in figure~\ref{fig:klfcomb} have not been
scaled in any way; the high redshift clusters are of similar richness as
Coma (although the comparison is not exact because the high redshift
LFs were obtained from a larger fraction of the
cluster area than for Coma; see below).
 The Coma
LF of \citet{and00} covers a smaller area than that of \citet{dep98}
but is in good agreement
(except for a dip in one bin). For the high
redshift clusters, the conversion from apparent to absolute magnitude was made using 
a \emph{k}-correction calculated for a 10Gyr old synthetic stellar population from the libraries of \citet{bru03}. The typical size of the \emph{k}-correction was $\approx-0.7$ mag.

\begin{figure}
\psfig{file=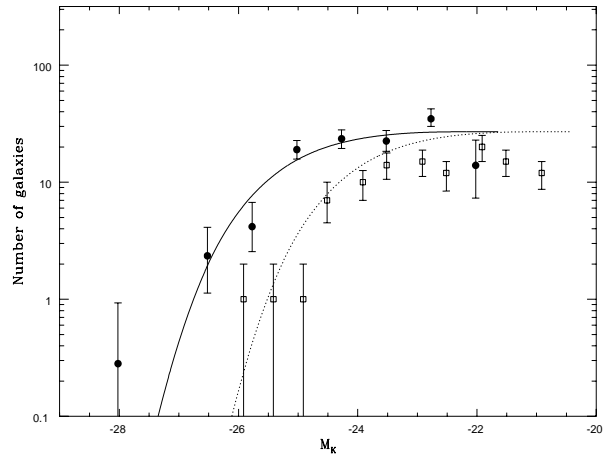,width=8.5cm,height=6.5cm,angle=270}
\caption{The combined luminosity function of three clusters including BCGs.  Symbols are as in figure~\ref{fig:klfcomb}.}
\label{fig:klfcomb_wbcg}
\end{figure}

The offset in absolute magnitude 
of $K\approx$1.2 mag between the Coma KLF and the combined high redshift KLF indicates the
degree of evolution. The same degree of evolution is also seen in figure~\ref{fig:klf_evol} as the difference 
between the observed values of $K^*$ and the no-evolution predictions. 
Evolving the fit to the high redshift KLF, assuming a passively
evolving model with  $z_{\rm{f}}=2$, gives a surprisingly good
description of the observed Coma KLF (the dotted lines in figures~\ref{fig:klfcomb} \&
\ref{fig:klfcomb_wbcg}). 
Thus differences between the high redshift and low redshift systems,
as shown by
the bright end of the LF studied here,  can be accounted for by
purely  passive evolution. There is no obvious change in the shape of
the LFs. 

\subsection{Integrated light functions.}
\label{sec:fracl}

An interesting and complementary measure of galaxy evolution within clusters is provided by their integrated light function.  This is calculated from the LF using

\begin{eqnarray}
\label{eqn:fracl}
{\sf j}(<L_{k}) = \frac{{\displaystyle \sum_{i=1}^{k}} N_{i} L_{i}}{{\displaystyle\sum_{i=1}^{n}} N_{i} L_{i}} & k=1, \ldots ,n 
\end{eqnarray}

where $N_{i}$ is the number of galaxies in the $i^{{\rm th}}$ bin of the LF and $L_{i}$ is the luminosity of the bin, and $n$ is the total number of bins.  For the BCGs and bins in which there is only one galaxy the individual contribution was calculated using the total luminosity of the galaxy and not the bin centre.

\begin{figure}
\psfig{file=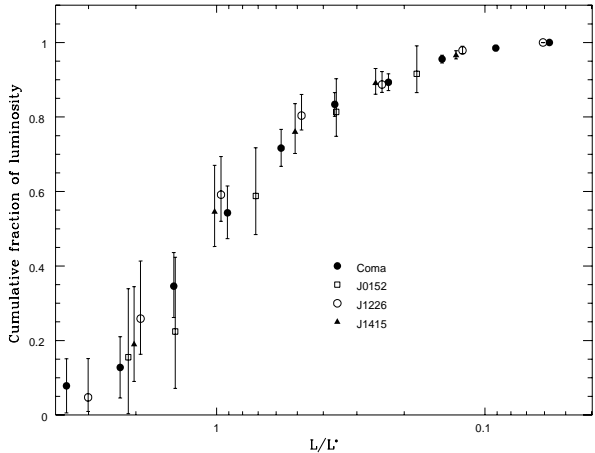,width=8.5cm,height=6.5cm,angle=270}
\caption{Integrated light functions for all three clusters compared to Coma.  Brightest cluster galaxies have been excluded.}
\label{fig:fracl}
\end{figure}

\begin{figure}
\psfig{file=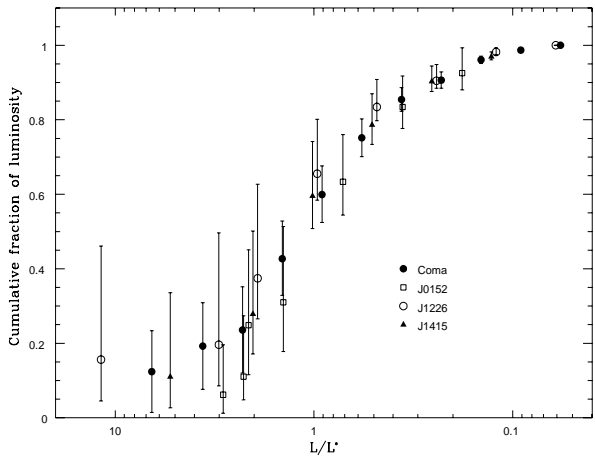,width=8.5cm,height=6.5cm,angle=270}
\caption{Integrated light functions as in figure \ref{fig:fracl} but with BCGs included.}
\label{fig:fracl_bcg}
\end{figure}

Integrated light functions were calculated for all three clusters and
for the Coma cluster.  These are shown in figures~\ref{fig:fracl} and
\ref{fig:fracl_bcg} for data excluding and including BCGs.  In order
to compare each cluster the luminosity has been normalised to $L^{*}$.
Thus the effects of any pure luminosity evolution (such as passive
evolution), which would alter the luminosity by a constant factor, are
eliminated from the plot and any variation between the shapes 
is due to some other evolutionary effect.  Note also that in order to
make this comparison the integrations of the light functions must
reach the same depth below $L^{*}$.  As this was not the case  the
counts for the faintest bin for ClJ1415 and the two faintest bins for
ClJ0152 were calculated from the best fitting Schechter functions and
thus these are not shown in the plots. The contribution from these
bins is only 5--10\% of the light integrated to our limit of 
$L/L^*=0.05$. Integrating
the Schechter function to luminosities fainter than this limit would
add only a further $\approx$ 1\%. 

A comparison of the shapes of the integrated light functions for the four clusters excluding their BCGs, shown in figure~\ref{fig:fracl}, shows 
that they are all consistent.  
If the BCGs are included in the integrated light function, as in figure~\ref{fig:fracl_bcg}, the light functions are generally consistent within errors but the scatter is greater.

If merging plays an important role in the evolution of \emph{massive}
galaxies since $z \approx 0.9$ then it would be expected that there
would be a higher fraction of total cluster light in bright galaxies
at low redshift compared to high redshift.  Little evidence is seen
for this in either figure~\ref{fig:fracl} or \ref{fig:fracl_bcg}.   If
BCGs are included then ClJ1226 is seen to have a slightly higher
fraction of light in its bright galaxies than Coma and ClJ0152 has a
lower fraction.
The lack of any systematic difference between the integrated light functions of the high redshift clusters and Coma  is suggestive that merging does not play a dominant role in the evolution of massive early-type galaxies  since $z \approx 0.9$.

One caveat is that interpretation of the integrated light
functions is complicated by the different areas observed for each
cluster.  The observations of the high redshift clusters reach radii
of 30\%--40\% of the virial radii (as measured from the X-ray
properties and provided 
by Ben Maughan, private communication), whereas the observation of the
Coma cluster covers only 0.16 of the virial radius ($r_{200}=2.3$ Mpc
for $H_{0}=70$km s$^{-1}$ Mpc$^{-1}$, \protect\citealt{rei02}).
\citet{dri98} show that the dwarf to giant galaxy ratio within rich
clusters is strongly anti-correlated with the mean projected density
outside the cores of clusters. Our observations do not, however, reach
dwarf galaxy luminosities. \citet{and01} confirms that the fraction of
dwarf galaxies increases in the outer parts of a $z=0.3$ cluster,
AC118, and furthermore shows that the faint end slope of the KLF
recovered from the outer parts of the cluster is considerably steeper
than the faint end slope of the KLF recovered from the core.
Consequently comparisons of observations covering different fractions
of the virial radius must be made with these effects in mind. However, our
concern is with the bright end of the LF in
the inner regions of the clusters, where these effects are unlikely to
produce large differences. 


\section{The brightest cluster galaxies.}
\label{sec:bcg}

The uneasy fitting of some BCGs on the Schechter function may be symptomatic of a different evolutionary process to the majority of galaxies within a cluster.  We now examine this evolution.

\begin{figure}
\psfig{file=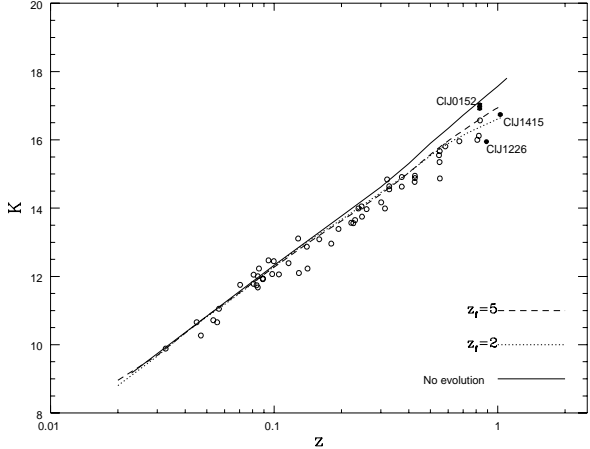,width=8.5cm,height=6.5cm,angle=270}
\caption{\emph{K} band Hubble diagram for BCGs.  Solid points are from
  this paper, open 
points are from \protect\citet{bro02}. Magnitudes are measured within
  a $12.5h^{-1}$ kpc radius aperture.}
\label{fig:khubble}
\end{figure}

The \emph{K} band Hubble diagram is shown in figure~\ref{fig:khubble}.
The models are the same as described above, only now they are
normalised to the low redshift ($z<0.1$), high $L_{\mathrm{X}}$ ($>
1.9 \times 10^{44}$ ergs s$^{-1}$ 0.3 -- 3.5 keV)  BCG photometry of
\citet{bro02}. The high $L_{\mathrm{X}}$ clusters from figure~1 of
\citet{bro02} are shown as the open circles.  The BCGs in the current
study, shown as solid points, extend to higher redshifts.  Note that
there are two BCGs shown for ClJ0152, one for each sub-clump.  The
magnitudes of the BCGs were measured in a metric aperture of radius,
$r_{\mathrm{m}}=12.5h^{-1}$kpc  for comparison with \citet{bro02}. The
BCG in ClJ1226 overlaps with at least one fainter galaxy observed in
projection; this galaxy has been separated by Sextractor. 
We note that the total magnitude of the ClJ1226 BCG measured here
($K$=15.34 mag) is consistent with the isophotal value of $K$=15.5 
measured by Cagnoni et al (2001).

It can be seen from figure~\ref{fig:khubble} that the BCGs in ClJ1226
and ClJ1415 are somewhat brighter than no-evolution predictions but in
good agreement with passive evolution models.  The BCGs of ClJ0152 are
fainter relative to the models than any other BCGs in
figure~\ref{fig:khubble} at $z>0.4$, perhaps indicative that the BCGs
are not fully formed.  Indeed as ClJ0152 is probably a merging system (\protect\citealt{mau03a}) it may be supposed that 
if the BCGs merge, and no further star-formation is triggered, they
will produce a brighter BCG more consistent with the other high
redshift BCGs.

In contrast ClJ1226 has a relaxed X-ray morphology
(\protect\citealt{mau03b}, \protect\citealt{ebe01}), and is 
one of the brightest BCGs at $z>0.4$ relative to the models. 
These results are suggestive that the evolution of BCGs at high
redshifts may be related
to the dynamical history of the cluster they inhabit, although
confirmation awaits a larger sample.

Table~\ref{tab:bcg_fracl} lists the contributions from the BCG to the
total $K$ band light for each cluster. The total luminosity in each
cluster was calculated from the LFs, as for the
integrated light functions, to a depth of $L/L^* \approx 0.06$.
The subclumps in ClJ0152 have been treated independently, since each
 has a BCG.  In the case
of Coma a correction was made to account for the smaller fraction of
the virial radius observed.  This was calculated using the number
density profile of \citet{ken82} to determine the ratio of the number
of galaxies within the observed area to that within $0.37r_{{\rm
    vir}}$, the average area observed for the high redshift
sample. The correction was a factor of 1.96.  The luminosity of the
Coma BCG, NGC4874, was calculated using the 2MASS total magnitude.  Note that there are two galaxies of similar
brightness in Coma, NGC4874 and NGC4884. 
We have taken NGC4874 as the BCG here as it resides at the centre of the X-ray emission.

\begin{table}
\caption{Fraction of the cluster $K$-band light  in the BCGs. The values for the BCGs in ClJ0152 are the fractions in each sub-clump independently. }
\label{tab:bcg_fracl}
\begin{tabular}{llll}
Cluster & Fraction & Radius/ $r_{\rm{vir}}$ & $K_{{\rm BCG}}^{Tot}$\\ \hline
ClJ0152A & 0.10 & 0.29 & 16.87\\
ClJ0152B & 0.12 & 0.34& 16.62\\
ClJ1226 & 0.16 & 0.41& 15.34\\
ClJ1415 & 0.11& 0.43& 16.27\\
Coma & 0.06 & 0.37$^{*}$& 8.86
\end{tabular}
\\
$^{*}$ extrapolated from r=0.17$r_{\rm{vir}}$, see text. 
\end{table}

The results show that the fraction of light in the BCGs at high
redshift is greater than, or equal to, the fraction in Coma.   
This is still true if both brightest galaxies in Coma are included.


There are uncertainties
 in the corrections applied in order to compare the clusters, which
 were originally measured to slightly different depths and covered
 different areas, although the variations in radius would produce only
 small variations in total $K$ band light ($\sim$20\% using the Coma
 light profile). A physical interpretation is that the high redshift BCGs are
 already of similar or greater mass to the BCG in Coma with no need to
 hypothesise any significant mergers in their future evolution. 

Why are the BCGs in ClJ0152 significantly fainter than the BCG in
ClJ1226? We will test the assumption that the stellar mass of the BCG is
related to the mass of its host cluster. We assume the ratio
of masses of the BCGs in ClJ0152 and ClJ1226 is proportional to the
ratio of the total cluster masses times the ratio of the fractions of
galaxy light in the BCGs,
assuming the mass fraction in galaxies is the same in each cluster.
Then the combined effects of the lower total mass
 of each subclump of ClJ0152 (by a factor of $\approx$0.39 compared to 
ClJ1226) and the lower fraction of cluster $K$-band light in the BCGs (by a factor
 of $\approx$0.69) are consistent with the difference in the
 luminosities of the BCGs (by 1.57 mag or a factor of 0.24).
This suggests that the BCG stellar masses at high redshift may be
 lower in lower mass host clusters, as inferred by \citet{bro02}, \citet{bur00} and \citet{nel02}.

We note that if the BCGs in ClJ0152 were to merge, as was hypothesised
 to explain their positions in the Hubble diagram, the fraction of
 light in the BCG would remain roughly similar.


\section{Discussion and conclusions}
\label{sec:dicuss}

The evolution of the galaxy populations of three high redshift
clusters of galaxies has been studied.  The bulk evolution of the
galaxies, as characterised by $K^{*}$, is found to be consistent with
passive evolution with a redshift of formation $z_{\mathrm{f}}\sim\
1.5$--2.  Further evidence for passive evolution is seen in the
similarity of the shape of the high-redshift LF with
that of Coma, and in the consistent shapes of the integrated light
functions. \citet{tre98b} also reached similar
conclusions about the evolution of the shape of the $K$-band
LF,
albeit with poorer accuracy.

Purely passive evolution of early-type galaxies is consistent with several other studies including the evolution of the $K$ band luminosity 
function (\protect\citealt{dep99}), evolution of the fundamental plane in terms of mass-light ratios (\protect\citealt{van98}), and studies of 
the scatter of the colour-magnitude relation 
(see e.g.\ \protect\citealt{ell97}, \protect\citealt{sta98}).
 
Our conclusions are different to those of \citet{bar98}, who
found no evidence of evolution in $M_K^*$ for clusters at
0.31$<$z$<$0.56 assuming a q$_0$=0.5 cosmology. However, the
cosmological dependence is such that at these redshifts the q$_0$=0.5
NE prediction is very similar to the ($\Omega_m$=0.3,
$\Omega_{\Lambda}$=0.7) $z_f=2$ passive evolution prediction (see
e.g. figure 8 of \protect\citealt{dep99}). Thus the results of Barger et
al. are  in agreement with those found here. 



When discussing formation it is important to distinguish between the epoch at which the stars in the galaxies were formed and the epoch at which the galaxies were assembled.  The studies of the fundamental plane and the colour-magnitude relation refer to the epoch of star formation.  If merging were a dissipationless process then it would be possible to have no extra star formation as a result of a merger and thus the age of the stars within a galaxy could be older than the age of galaxy assembly.   A study of the cluster of galaxies MS 1054-03 at $z=0.83$ is presented by \citet{van99} and \citet{van00} in which there is observed a high fraction of merging red galaxies.  Very little star formation is seen in the merging galaxies constituting evidence that the galaxies are in fact somewhat younger than the stars that reside within them.  

Is such merging reflected in the evolution of the LF?
The $K$ band luminosity of a galaxy is very nearly independent of
star-formation, but reflects the mass of the old stars within the
galaxy.  Thus $K$ magnitudes are a good measure of the stellar mass of
a galaxy.  \citet{dia01} give predictions of the evolution of the KLF
from semi-analytic models of dissipationless, hierarchical structure
formation.  The models show that there is very little evolution of the
number of massive galaxies in clusters since $z=0.8$.  The galaxies
are assembled at high redshift and evolve passively with little
subsequent merging after $z=0.8$. 
A detailed comparison with these models is however not possible
because the  luminosity evolution predictions in the models are not
sufficiently accurate for massive galaxies (\protect\citealt{dia01}).

It is clear from figure \ref{fig:klf_evol} that the bulk of
galaxies in our sample were brighter in the past than predicted from
no-evolution models, by $\Delta K \approx$ -1.2 mag at $z=0.9$.  
A direct comparison of the high-redshift $K$ band LF
with that of Coma suggests that the two are very similar in shape and
the differences in $M_{K}^{*}$ may be reconciled by pure luminosity
evolution. The fading with time by $\Delta K \approx$ -1.2 mag is consistent
with passive evolution from a  formation epoch $z_{f}\approx 2$.
In the
monolithic collapse picture this would be expected naturally.   The models of \citet{dia01} show
that this is also predicted for massive cluster galaxies in a hierarchical scenario  since most merging takes place early on the history of the cluster.
  
In a merging model passive evolutionary processes will still be
present, and thus  $K^{*}$ would still appear brighter than
no-evolution predictions.  A probe of `extrapassive' processes is the
shape of the LF. 
It is found that $\alpha$ is consistent with that of Coma  although it is poorly constrained here.  
Perhaps a stronger test for extrapassive processes is the shape of the
integrated light function. 
 The lack of any major changes seen in figure~\ref{fig:fracl} is suggestive that passive evolution alone is responsible for the evolution measured in $K^{*}$. We conclude that the luminosity evolution of bright galaxies in massive clusters is consistent with pure passive evolution, but note that this may be consistent with hierarchical models if most merging takes place at high redshifts.


The mild $K$-band evolution of luminous field galaxies to z=1 found in
recent surveys is also consistent with luminosity evolution, although
the amount of evolution  $\Delta M_K^*=-0.54\pm0.12$ mag
(\protect\citealt{poz03}) or $\Delta M_K=-0.7\pm0.3$ mag (\protect\citealt{feu03}, \protect\citealt{dro03}, see also \protect\citealt{im02}) is less than that
observed here ($\Delta M_K^*=-1.2\pm0.3$ mag) or by \citet{dep99}
($\Delta M_K^*=-0.90\pm0.25$) in massive clusters.
The recent results on the field evolution indicate 
little change in
$\phi^*$ to z=1, and are in contrast to the conclusions drawn by 
\citet{kau98b} and
 \citet{kau96} who found evidence for a deficit of massive galaxies in
 the field at $z\approx 1$. 

The difference in luminosity evolution between the
 field and massive clusters (both at z$\approx$0.9) is in the opposite 
sense to the environmental dependence of the formation epoch in hierarchical models (\protect\citealt{kau93}, \protect\citealt{bau96},
 \protect\citealt{dia01}). In these models, the assembly of an
 early-type galaxy, and the formation of its stars, occurs at an
 earlier epoch in a region of high overdensity than in the field. The degree of
 luminosity evolution  between z=1 and z=0 is predicted, therefore, to
 be higher in early-type field galaxies than in massive clusters, since  z=1  is closer to
 their epoch of formation. If the high redshift measurements are not
 in error (they can be checked via spectroscopic samples), then more
 detailed studies will be required to separate the $K$-band evolution
 of galaxies of different  spectral types, 
since  the cluster and field samples contain a different mix of
 galaxies and thus a different mix of star formation histories.


Note that although there is a strong case that the observed evolution
in $K^{*}$ may be attributed almost entirely to passive evolutionary
processes, the interpretation of this result as being due to a
redshift of formation $z_{f}=1.5-2$ is less secure.  The models used
are for a single burst of star-formation for a stellar population with
a Salpeter initial mass function having a solar
metallicity. Age-redshift relations are calculated for an assumed flat
cosmology with $H_{0}=70$km s$^{-1}$ Mpc$^{-1}$ and
$\Omega_{\rm{M}}=0.3$ and $\Omega_{\Lambda}=0.7$.  All of these
assumptions affect the resultant redshift of formation and so there is
clearly some slack in the interpretation of the evolution.  For example if the models if figure~\ref{fig:klf_evol} are normalised to the high $L_{{\rm X}}$ point at $z=0.2$, rather than Coma, the data are more consistent with $z_{{\rm f}}=2$--5. Future
work based on the colours and morphologies of the member galaxies will
provide stronger constraints on the epoch of formation. 


The evolution of BCGs is consistent with that found by \citet{bro02}
for high $L_X$ clusters.  Figure \ref{fig:khubble} exhibits a degree of scatter in the evolution of BCGs at high redshift with some BCGs being consistent with no-evolution predictions, of which ClJ0152 is an example, and others being consistent with passive evolution.  This can be interpreted as showing that some BCGs are fully formed at high redshift e.g.\ ClJ1226, whereas others would need to undergo merging between $z=1$ and the present to reconcile them with local BCGs. 

This is supported by the ratios of $K$ band light in the BCGs to total
cluster light.  The ratio in Coma is not larger than the ratio in the high
redshift clusters. Thus excess brightening of the BCGs with time,
relative to the other cluster galaxies, as would be
expected from such processes as cannibalism, is not observed,  
suggesting that the BCGs, at least  of ClJ1226 and ClJ1415, are already fully formed.



\section*{Acknowledgments.}

The authors would like to thank warmly both Ben Maughan, for his role
in observing at UKIRT 
and discussions about the X-ray properties of the clusters, and Harald
Ebeling, for providing all the galaxy redshifts. UKIRT staff have been
very efficient and helpful; service time
observations were performed by some of them.  
The authors would also like to thank 
Sarah Brough for providing the
data for figure~\ref{fig:khubble}, Caleb Scharf for suggestions based
on an earlier version of this paper, and 
Stefano Andreon and Antonaldo Diaferio for their help.  The authors would like to thank the referee for helpful comments which have improved this manuscript. SCE
acknowledges a PPARC studentship. This research has made use of the NASA/ IPAC Infrared
                                                                      Science Archive, which is operated by the Jet Propulsion
                                                                      Laboratory, California Institute of Technology, under
                                                                      contract with the National Aeronautics and Space
                                                                      Administration.

\bibliographystyle{mn2e}
\bibliography{clusters}

\end{document}